\documentclass[12pt]{iopart}

\usepackage{iopams}
\usepackage[compress]{cite}
\usepackage{graphicx}
\usepackage{color} 
\begin{document}

\title[]{Extended model for optimizing high-order harmonic generation in absorbing gases}

\author{Bal\'azs Major, Katalin Varj\'u}
\address{ELI-ALPS, ELI-HU Non-Profit Ltd., Wolfgang Sandner utca 3., Szeged 6728, Hungary}
\address{Department of Optics and Quantum Electronics, University of Szeged, D\'om t\'er 9., Szeged 6720, Hungary}
\ead{Balazs.Major@eli-alps.hu}

\vspace{10pt}
\begin{indented}
\item[]2021
\end{indented}

\begin{abstract}
We report on an extended version of the one-dimensional model proposed by Constant \emph{et al.} [Phys. Rev. Lett 82(8), 1668 (1999)] to study phase matching of high-order harmonic generation in absorbing and dispersive medium. The model  --- expanded from zeroth to first order --- can be used with media having a pressure profile varying linearly with propagation length. Based on the new formulas, the importance of having a generation medium that ends abruptly with a steep pressure gradient for achieving high flux is highlighted. In addition to further rule-of-thumb guidelines for harmonic-flux optimization, it is shown that having a steep increase of pressure in the beginning of the medium increases harmonic flux, while it also decreases the required medium length to reach the absorption-limited maximum.
\end{abstract}

%
%
%
%
%

\section{Introduction}

Since its first demonstration in the end of the 1980's \cite{Ferray1988JPB, McPherson1987JOSAB}, high-harmonic generation (HHG) in atomic gases has become the most widely used method to generate ultrashort pulses in the extreme ultraviolet (XUV) and X-ray wavelength regimes \cite{Chang2016JOSAB, Heyl2016JPB, Calegari2016JPB, Kuhn2017JPB}. These attosecond-pulse sources are used for a wide range of applications with exponentially growing coverage of research fields \cite{Li2020NatComm}. This includes a vast number of topics in chemistry and biology \cite{Lepine2014NPhot, Leone2016FarDis, Nisoli2017ChemRev}; in material science of solids \cite{Ghimire2014JPB, Sederberg2020NatComm} and liquid phase samples \cite{Jordan2020Science}; or in different areas of quantum \cite{Shi2020JPB}, atomic \cite{Isinger2020Science}, molecular \cite{Biswas2020NatPhys} and nonlinear physics \cite{Orfanos2020JPhysPhot}. 

One drawback of gas-target HHG is the low generation efficiency \cite{Tate2007PRL, Heyl2016JPB}, so there is continuous effort to increase performance and to scale up the power of HHG beamlines \cite{Sansone2011NatPhot, Heyl2016JPB, Heyl2016Optica, Major2021Optica}. The macroscopic generation process (the interaction of multiple atoms with the intense laser field) is a very complex mechanism involving phase matching as well as reabsorption \cite{Gaarde2008JPB, Heyl2016JPB}, well-known in the field of nonlinear optics \cite{Boyd}. 
In the past decades an extensive literature has formed focusing on the complex aspects of phase matching in HHG. This includes several comprehensive theoretical works (often applying numerical methods) \cite{Ruchon2008NJP, Dachraoui2009, Wang2011PRA, Kazamias2011PRA, Rogers2012JOSAB} and thorough experimental investigations \cite{Popmintchev2009PNAS, Rothhardt2014NJP, Sun2017Optica, Schotz2020PRX}, along with a vast number of tutorials and reviews \cite{Gaarde2008JPB, Heyl2016JPB, Hareli2020JPB}. 

Nevertheless, with simplifying assumptions certain simple rules can be identified to optimize \cite{Constant1999PRL, Heyl2016JPB}, power-scale \cite{Heyl2016Optica} or intensity-scale high-harmonic sources \cite{Senfftleben2020JPhysPhot}. 
The first set of thumb rules was established more than twenty years ago by Constant \emph{et al.} \cite{Constant1999PRL} --- followed by the independent demonstration of experimental applicability shortly after \cite{Tamaki2000PRA} ---, and is still followed nowadays when designing and optimizing HHG beamlines \cite{Porat2018NPhot, Klas2020OE, Pupeikis2020Optica, Ye2020JPB}. This analytical, one-dimensional model simplifies the description of the macroscopic HHG process where all parameters (atomic number density, single-atom response, phase mismatch, absorption) related to the harmonic build-up are constant along the generation medium. Here, we move a step forward, and extend this model to situations where most of these parameters vary linearly with propagation distance in the medium. This allows us to study the HHG process under more realistic conditions, but still in an analytical manner; providing general, rule-of-thumb optimization guidelines. We use the resulting formulas to study how pressure gradients at the beginning and end of the gaseous medium affect the achievable XUV flux.

The present work is structured as follows: in Section \ref{sec:Methods}, we revise the model of Constant \emph{et al.} \cite{Constant1999PRL} to introduce the methods used in this paper. In Section \ref{sec:Results}, first we shortly discuss the physical considerations allowing us to extend the model for non-static pressure generation media. Then we derive expressions that are used to study the effect of linear pressure gradients at the beginning and end of the generation volume on the harmonic build up. In Section \ref{sec:Summary}, we summarize our main conclusions.

\section{Methods}\label{sec:Methods}

\subsection{One-dimensional model of phase matching in absorbing gases}

As is known from textbooks on nonlinear optics \cite{Boyd}, a certain harmonic order $q$ of the fundamental wave must obey the nonlinear wave equation, which is an inhomogeneous partial differential equation. According to the  solution of this equation, intensity of the field oscillating with angular frequency $\omega_{q}$ will depend on the value of
\begin{equation}\label{eq:pmismatch}
	\Delta k = q k_{l} - k_{q}\,,
\end{equation}
$k_{q}$ being the wave number of the field of harmonic order $q$, and $k_{l}$ meaning the same property of the fundamental, generating field. Eq. (\ref{eq:pmismatch}) represents the wave vector mismatch between the $q^{\mathrm{th}}$ harmonic field and the induced polarization at the $q\omega_{l}$ frequency. This approach can be generalized for high harmonic orders \cite{Heyl2016JPB}. A more illustrative quantity for the amount of phase mismatch  is the coherence length $L_{\mathrm{coh}} = \pi/\Delta k$, defined as the propagation length in which the radiation constructively builds up.

At the same time, it is not just this phase difference that determines the harmonic intensity, but the generated $\omega_{q}$ photons are also absorbed in the medium. This absorption can be characterized by the absorption coefficient $\kappa$, leading to an exponential decay of the field amplitude during propagation along axis $z$ \cite{Hecht}. Similarly to the coherence length, an absorption length $L_{abs} = 1/(2\kappa)$ can be used to quantify the strength of absorption, introduced as the propagation distance after which the intensity is decreased to $1/e$ times its original value \cite{Heyl2016JPB, Constant1999PRL}.  

In general, the flux of the $q^{\mathrm{th}}$ harmonic field on-axis can be calculated as \cite{Constant1999PRL, Kazamias2003PRL, Heyl2016JPB}
\begin{equation}\label{eq:integral}
	S_{q} \propto |E_{q}|^2 = \left|  \int_{0}^{L_{\mathrm{med}}}  A \rho(z)
	\exp \left( \mathrm{i} \left[\Delta k(z) + \mathrm{i} \kappa(z)\right]
	\left[L_{\mathrm{med}}-z\right] \right) \mathrm{d}z
	\right|^{2} \,,
\end{equation}
where $A$ is the strength of the generated field (amplitude of single-atom response), $\rho(z)$ is the number density of atoms, and $L_{\mathrm{med}}$ is the length of the generation medium. Quantities $A$, $\Delta k$ and $\kappa$ depend on the harmonic order $q$, but indication of this dependence is omitted here for brevity. 

\begin{figure}[htb!]
	\begin{center}
		\includegraphics[width=8cm]{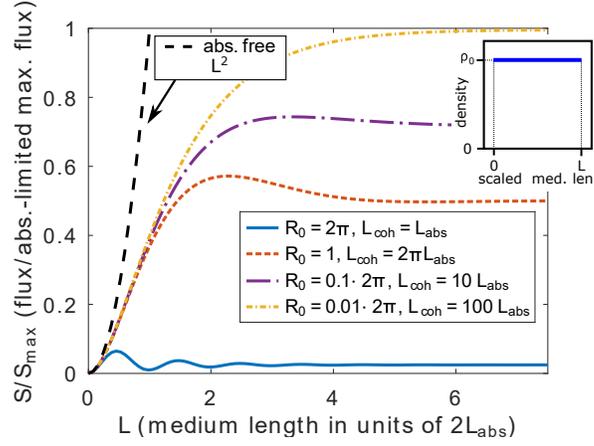}
	\end{center}
	\caption{\label{fig:Constant}The high-harmonic flux $S$ (normalized to the absorption-limited maximum $S_{\mathrm{max}}$ of Eq. (\ref{eq:Smax})) as a function of medium length generated in a medium with constant pressure profile along the laser beam propagation axis. Inspired by Refs. \cite{Constant1999PRL, Heyl2016JPB}. The different curves show the flux for certain ratios $R_{0} = \Delta k_{0} / \kappa_{0} = 2 \pi L_{\mathrm{abs}} / L_{\mathrm{coh}}$ of the absorption and coherence lengths. The same medium length range is plotted as in FIG. 1 of \cite{Constant1999PRL}. The inset shows the medium density/pressure profile assumed. The black dashed curve shows the absorption-free, perfectly phase-matched case, represented by an $L^{2}$ curve in this plot.}
\end{figure}

Assuming a constant pressure\footnote{Atomic number density and pressure are used as synonyms in this work, assuming a constant temperature, $\rho$ representing a quantity of $\mathrm{1/m^3}$ dimension in every occurrence.} profile $\rho(z) = \rho_{0} = 2 \kappa_{0}/\sigma$ ($\sigma$ being the photoionization cross section \cite{Constant1999PRL}), a $z$-independent dipole amplitude $A$, along with constant phase mismatch $\Delta k(z) = \Delta k_{0}$ ($>0$\footnote{Since sometimes phase mismatch is defined with opposite sign \cite{Heyl2016JPB}, $\Delta k_{0}>0$ does not mean any physical restriction.}) and absorption $\kappa(z) = \kappa_{0}$ ($>0$), the harmonic flux $S_{q}$ for a certain harmonic order $q$ can be analytically evaluated  for both non-guiding \cite{Constant1999PRL, Heyl2016JPB} and guiding generation geometries \cite{Durfee1999PRL} (see also \ref{Asec:zeroth}). The resulting 
formula (equivalent to the expressions introduced by Heyl~\emph{et.\,al.} \cite{Heyl2016JPB} and Constant~\emph{et.\,al.} \cite{Constant1999PRL}) is

\begin{eqnarray}\label{eq:Heyl_dimless}
	|E_{q}|^2 &= \frac{8 A^{2}}{\sigma^{2}} 
	\exp\left(-L\right)
	\frac{\cosh (L) - \cos(R_{0} L)}{1 + R_{0}^{2}} \\
	&= \frac{4 A^{2}}{\sigma^{2}}
	\frac{1}{1+R_{0}^2} 
	\left[
	1 + \exp \left( - 2 L \right)
	- 2 \cos \left( R_{0} L \right)
	\exp \left( -L \right)
	\right]
	\,, \label{eq:Constant_dimless}
\end{eqnarray}
where $L = \kappa_{0} L_{\mathrm{med}} = L_{\mathrm{med}}/(2 L_{\mathrm{abs}})$ is the dimensionless measure of the medium length and $R_{0} = \Delta k_{0} / \kappa_{0} = 2 \pi L_{\mathrm{abs}} / L_{\mathrm{coh}}$ is another dimensionless variable quantifying relative strength of phase mismatch and absorption. The advantage of Eqs. (\ref{eq:Heyl_dimless}) and (\ref{eq:Constant_dimless}) is that due to their dimensionless forms they can serve general and universal guidelines for optimizing phase matching in static-pressure, dispersive and absorptive media. The high-harmonic flux for different $L$ and $R_{0}$ values is depicted in Figure \ref{fig:Constant}.

It is easy to see from Eq. (\ref{eq:Constant_dimless}) that the limit at infinite medium length $L \rightarrow \infty$ ($L_{\mathrm{med}} \rightarrow \infty$) is  
\begin{equation}\label{eq:lim_infty}
	S_{\infty} = \frac{4 A^{2}}{\sigma^{2}}
	\frac{1}{1+R_{0}^2} \,.
\end{equation}
The above mathematical limit of Eq. (\ref{eq:lim_infty}) means two important physical implications on the achievable flux in absorptive media. First, the absorption-limited flux even with perfect phase matching ($R_{0} \rightarrow 0$) has a maximum of
\begin{equation}\label{eq:Smax}
S_{\mathrm{max}} = \frac{4 A^2}{\sigma^2}\,, 
\end{equation}
compared to the limitless value of the absorption-free and perfectly phase matched case of $S_{\mathrm{abs-free}}/S_{\mathrm{max}} = L^2$ (when $\kappa_{0} = \Delta k_{0} = 0$), represented by the black dashed curve in Figure \ref{fig:Constant}. Second, the flux achievable at $L_{\mathrm{med}} \rightarrow \infty$ depends only on the value of $R_{0}$, so the ratio of coherence  ($L_{\mathrm{coh}}$) and absorption ($L_{\mathrm{abs}}$) lengths.
This leads to the well know rule-of-thumb optimization conditions (assuming an unlimited medium length is realizable) of 
\begin{eqnarray}
R_{0} < 1 \quad ( \Rightarrow L_{\mathrm{coh}} > 2 \pi L_{\mathrm{abs}}) \quad \mathrm{and} \label{eq:Rgt1} \\
L > 1.5 \quad (\Rightarrow L_{\mathrm{med}} > 3 L_{\mathrm{abs}})\,, \label{eq:Lgt1p5}
\end{eqnarray}
giving at least half of the maximum signal achievable with absorption-limited generation \cite{Constant1999PRL, Heyl2016JPB}. Another important result highlighted in Ref. \cite{Constant1999PRL} is that the achievable flux is independent of pressure\footnote{It is to be noted that from the physical point of view the flux of course indirectly depends on the pressure through $L_{\mathrm{abs}} = 1/(\sigma \rho)$.} (if $\rho_{0} = 2\kappa_{0}/\sigma >0$), and optimization of flux can be obtained by simultaneous increase of $A^{2}/\sigma^{2}$ (basically maximizing the single-atom response $A$) and fulfillment of the above requirements  on $L_{\mathrm{coh}}$ and $L_{\mathrm{med}}$ (Eqs. (\ref{eq:Rgt1}) and (\ref{eq:Lgt1p5})).

\section{Results and discussion}\label{sec:Results}

In the following, we analytically extend the model described in Section \ref{sec:Methods} by assuming linear variation of pressure ($\rho(z) = \rho_{0} + \rho_{1}z$), phase mismatch ($\Delta k(z) = \Delta k_{0} + \Delta k_{1}z$) and absorption ($\kappa(z) = \kappa_{0} + \kappa_{1}z$) as a function of propagation distance $z$, while keeping the presumption of a constant single-atom response $A$\footnote{We note here that analytical solution can also be obtained when assuming a linear dependence of $A$ with $z$, in addition. However, differently from the cases of $\kappa$ and $\Delta k$ --- described in Section \ref{subsec:dk} ---, this form of variation for the single atom response $A$ is physically not justifiable.}. 
Mathematically, the assumption of linear dependence is equivalent to Taylor series expansion of the above quantities up to first order in $z$. In this sense the original model of Constant \emph{et al.} \cite{Constant1999PRL} can be considered as zeroth order, while the following expressions are special cases of the first-order approximation. Before deriving the new formulas, we analyze when the above mentioned linear dependencies are physically meaningful. Detailed explanations and derivation steps of formulas presented in later subsections can be found in the Appendix.

\subsection{Phase matching terms and their pressure dependence in high-harmonic generation}\label{subsec:dk}

Assuming a pressure profile of the form $\rho(z) = \rho_{0} + \rho_{1}z$ directly gives a similar dependence of absorption $\kappa(z)$ on $z$ through the relation $\rho(z) = 2 \kappa(z)/\sigma$ between pressure and absorption. A similar, linear form of phase mismatch evolution, however, is not general. So first, conditions under which phase mismatch can be written in the form of $\Delta k(z) = \Delta k_{0} + \Delta k_{1}z$ has to be discussed.

Using the nomenclature of Refs. \cite{Heyl2016JPB, HeylPhD}, phase mismatch can be written as a sum of the following four terms:
\begin{equation}\label{eq:dk4}
	\Delta k = \Delta k_{g} + \Delta k_{d} + \Delta k_{n} + \Delta k_{p}\,,
\end{equation}
where $\Delta k_{g}$ is a geometrical term related to the spatial phase variation of the generating laser beam, $\Delta k_{d}$ is the phase mismatch induced by the atomic dipole phase, $\Delta k_{n}$ is caused by dispersion in the atomic medium, while $\Delta k_{p}$ is the plasma term resulting from the presence of free electrons (originating from ionization of atoms during the generation process). In the following paragraphs, each of the four terms is analyzed from the aspect of direct or indirect (through pressure $\rho(z)$) dependence on $z$.

In general, the geometrical phase term $\Delta k_{g}$ changes non-linearly with $z$ \cite{Heyl2016JPB}. However, according to the considerations in Ref. \cite{HeylPhD}, the geometrical phase mismatch term is constant for generation in a capillary \cite{Durfee1999PRL}, or in a focusing geometry where the medium length $L_{\mathrm{med}}$ is similar to or smaller than the Rayleigh length $z_{R}$ of the laser beam ($L_{\mathrm{med}} \lesssim z_{R}$). The term is also small very far from the laser focus, where on-axis phase variations of the fundamental field are slow \cite{Major2015JO}.

Similarly, the dipole-induced phase mismatch has a complex spatial dependence through the variation of laser intensity with $z$ ($\Delta k_{d} = \alpha \partial I_{0}/\partial z$ \cite{Lewenstein1995PRA}). At the same time, it can be shown that for a guiding capillary \cite{Durfee1999PRL}, in case of a self-guided beam \cite{Rivas2018Optica, Major2021Optica}, or in experiments where loose focusing geometry is used (expressed as $L_{\mathrm{med}} \ll z_{R}$ in a more quantitative form), this intensity variation is slow and the related phase-mismatch term can be neglected \cite{Heyl2016JPB}. Also, when only short quantum trajectories are relevant, the proportionality constant $\alpha$ is small, and $\Delta k_{d} \approx 0$ \cite{Heyl2016JPB}.

It is also easy to see that at typical pressures ($<2\,\mathrm{bar}$) and photon energies ($>1\,\mathrm{eV}$, including fundamental) involved in HHG, dispersion-related phase mismatch terms ($\Delta k_{n}$ and $\Delta k_{p}$) depend linearly on pressure. Using well-known formulas and Taylor expansions of elementary functions, the neutral dispersion gives \cite{Heyl2016JPB}
\begin{equation}\label{eq:dkn}
	\Delta k_{n} = \frac{\omega_{q}}{c}  \left( \frac{ \alpha_{\mathrm{dip}}}{2 \varepsilon_{0}} + \frac{r_{a} 2\pi c^2}{\omega_{q}^{2}}  f_{1}  \right) \rho = \gamma_{n} \rho \,,
\end{equation}
where $\alpha_{\mathrm{dip}}$ is the static polarizability of the constituting atoms, $r_{a}$ is the classical Bohr radius, $\omega_{q}$ is the angular frequency of harmonic order $q$, $c$ is the speed of light in vacuum, and $f_{1}$ is an atomic form factor \cite{Henke1993ADNDT}. In a similar manner, the plasma-related phase mismatch is
\begin{equation}\label{eq:dkp}
	\Delta k_{p} = -\frac{\omega_{q}}{c} \frac{e^2 \Gamma}{\varepsilon_{0} m_{e}}  \left( \frac{1}{\omega_{l}^{2}} - \frac{1}{\omega_{q}^{2}} \right) \rho = \gamma_{p} \rho\,,
\end{equation}
where $e$ is the elementary charge, $m_{e}$ is the mass of a stationary electron, $\Gamma$ is the ionization ratio and $\omega_{l}$ is the angular frequency of the laser field  \cite{Heyl2016JPB}. It is to be noted that while both (\ref{eq:dkn}) and (\ref{eq:dkp}) have a limited atom density/pressure range in which they can be used, only the second, Eq. (\ref{eq:dkp}) sets a limit on the photon energy range where the approximations are valid. 

The above altogether mean that in the described conditions --- which are most often fulfilled in HHG beamlines \cite{Heyl2016JPB}  --- phase mismatch contributions can be grouped to terms which are either independent of $z$, written as $\Delta k^{(c)} = \Delta k_{g} + \Delta k_{d}$, or related to atom number density $\rho$ through a proportionality constant ($\gamma_{n} + \gamma_{p}$), expressed here as $\Delta k ^{(\rho)} = \Delta k_{n} + \Delta k_{p}$. So, in the cases analyzed earlier by Constant \emph{et al.} \cite{Constant1999PRL}, the fixed values of absorption and phase mismatch are direct consequences of a constant pressure medium. Generally, in the above conditions, the change of atom density along the propagation axis $z$ will define the $z$-variation of absorption and phase mismatch\footnote{This means that the cases with non-linear dependence on $z$ can also be treated in a similar way as shown in the following. However, higher than first-order Taylor polynomials are not leading to analytical expressions, and such solutions lie outside the scope of this work.}.
This way in a generation volume with linearly varying pressure phase mismatch can be written as
\begin{eqnarray}
	\Delta k =  \Delta k^{(c)} + \frac{\partial \Delta k ^{(\rho)}}{\partial\rho}\rho = \Delta k^{(c)} +  (\gamma_{n} + \gamma_{p})(\rho_{0} + \rho_{1} z)\,,
\end{eqnarray}
and it is easy to see that this is of the form
\begin{equation}
	\Delta k(z) = \Delta k_{0} + \Delta k_{1} z\,,
\end{equation} 
with 
\begin{eqnarray}
	\Delta k_{0} = \Delta k^{(c)} + \frac{2(\gamma_{n} + \gamma_{p})\kappa_{0}}{\sigma}   \quad  \mathrm{and} \label{eq:dk0} \\
	\Delta k_{1} =  \frac{2(\gamma_{n} + \gamma_{p}) \kappa_{1}}{\sigma}\,. \label{eq:dk1}
\end{eqnarray}

Since the magnitude of these quantities determine the harmonic build-up process, typical values are given in \ref{Asec:dk_values}. In summary, the main point is that in practically relevant cases, phase mismatch and absorption show a wider range of variation than what is plotted in Figure \ref{fig:Constant}, or in later figures. Parameters in plots are chosen to show the range in which they have relevant effect on the high-harmonic flux.

\subsection{Phase matching in medium with linearly increasing pressure}\label{subsec:rho0zero}
\begin{figure}[htb!]
	\begin{center}
		\includegraphics[width=16cm]{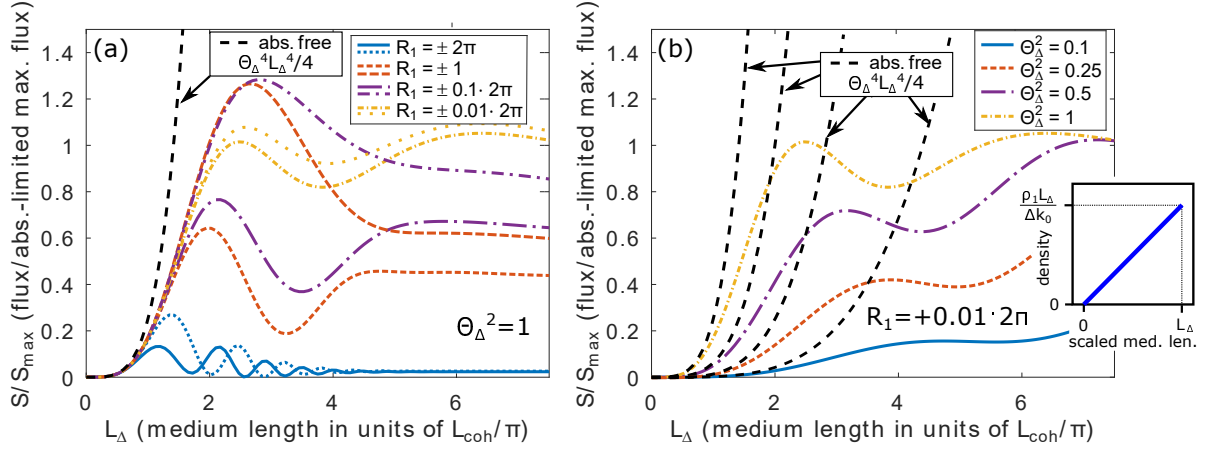}
	\end{center}
	\caption{\label{fig:rho0zero}The harmonic flux $S$ (normalized to the absorption-limited maximum $S_{\mathrm{max}}$ of Eq. (\ref{eq:Smax})) as a function of medium length $L_{\Delta} = \Delta k_{0} L_{\mathrm{med}}$. The medium has a linearly increasing pressure profile along the laser propagation direction (see inset of (b)). (a) Dependence of flux on ratio $R_{1}$ between the gradients of phase mismatch and absorption. (b) The same as a function of gradient steepness quantified by $\Theta_{\Delta}^2$. The black dashed curves show the absorption-free, perfectly phase matched case, depending on the value of $\Theta_{\Delta}^{2}$.}
\end{figure}

Now, let's assume a medium that can be described with a linear increase of pressure from zero, that is of the form $\rho(z) = \rho_{1} z$ with $ \rho_{1} > 0$ (see inset of Figure \ref{fig:rho0zero}). During HHG, usually the generating laser beam encounters a certain pressure gradient as it enters the generation volume, regardless if it is a gas cell \cite{Major2020JPysPhot}, (supersonic) jet \cite{Major2021Optica} or a gas-filled capillary/fiber \cite{Goh2015OE}. The linear increase of pressure at the beginning of the medium is the simplest form to assume. As described in \ref{Asec:first}, in this case evaluation of the integral in Eq. (\ref{eq:integral}) leads to:

\begin{eqnarray}\label{eq:rho0zero}
	E_{q} = \frac{A}{\sigma} \frac{1}{1- i R_{1}} \times \nonumber \\
	 \left\{ \sqrt{\pi}\left[ f_{\Delta} + g_{\Delta} \right]
		\exp{\left( - f_{\Delta}^{2} \right)} \right.
		\left[ \mathrm{erfi} \left( f_{\Delta} \right)
		+ \mathrm{erfi} \left(  f_{\Delta} + g_{\Delta} \right)  \right] + \nonumber \\
		 \left. \left[ 1 - \exp\left( g_{\Delta}^2 + 2 g_{\Delta} f_{\Delta} \right)  \right] \right\}
\end{eqnarray}
with a dependence on medium length $L_{\Delta} = \Delta k_{0} L_{\mathrm{med}}$ purely through
\begin{equation}
	f_{\Delta}(L_{\Delta}) = -\frac{g_{\Delta}}{2} \left[ i \Theta_{\Delta}^{2}(1-i R_{1}) L_{\Delta} + 1 \right]\,,
\end{equation}
and $R_{1} = \Delta k_{1}/\kappa_{1}$, $\Theta_{\Delta}^{2} = \kappa_{1}/\Delta k_{0}^{2}$ ($>0 \Leftarrow \kappa_1 > 0 \Leftarrow \rho_{1}>0$), $g_{\Delta} = i/(\Theta_{\Delta} \sqrt{1-i R_{1}}) $. Since $g_{\Delta}$ and $f_{\Delta}(L_{\Delta})$ quantities have only been introduced for brevity, Eq. (\ref{eq:rho0zero}) depends only on three dimensionless variables: $L_{\Delta}$, $R_{1}$ and $\Theta_{\Delta}^{2}$. Similarly to the constant pressure case (Eqs. (\ref{eq:Heyl_dimless}) and (\ref{eq:Constant_dimless})), two of these variables define the medium length ($L_{\Delta}$) and the ratio of phase-mismatch and absorption gradients ($R_{1}$). These, however, are defined with a different normalization variable ($\Delta k_{0}$ instead of $\kappa_0$), since due to zero pressure at the medium beginning, absorption is zero ($\kappa_0 = 0$), and it cannot serve as a normalization variable. The third dimensionless variable is $\Theta_{\Delta}^{2}$, quantifying the relation of first and zeroth order coefficients, so the steepness of the pressure gradient. Since Eq. (\ref{eq:rho0zero}) depends on three dimensionless variables, general rules can be set --- similarly to the zeroth-order model --- for optimizing phase matching. The plots serving as the basis of this analysis are summarized in Figure \ref{fig:rho0zero}. When using these expressions and plots for flux optimization in practice, one has to be careful that the dimensionless variables ($L_{\Delta}$, $R_{1}$ and $\Theta_{\Delta}^{2}$) depend on multiple physical quantities. So when  analyzing the effect of phase mismatch $\Delta k_0$, for example, both the scaled medium length $L_{\Delta}$ and the gradient factor $\Theta_{\Delta}^2$ change.    

For physically meaningful results, the analysis is restricted to $\Theta_{\Delta}^{2} > 0$ ($\Leftarrow \rho_{1} > 0$, meaning no negative pressure). While the integral $E_{q}$ can be evaluated analytically using tabulated integrals \cite{gradshteyn2007} (see details in \ref{Asec:first}), analytical evaluation of its modulus square is not possible due to the presence of the special function $\mathrm{erfi}(z) = -i \mathrm{erf}(iz)$, the imaginary error function \cite{gradshteyn2007}.

In the special case of infinite medium length ($L_{\Delta} \rightarrow \infty$) it is possible to evaluate $S_{q} = |E_{q}|^{2}$ analytically using Eq. (\ref{eq:rho0zero}) (see details in \ref{Asec:first}), giving the exact same value as for the constant pressure case (that is, Eq. (\ref{eq:lim_infty})). This means that $R_{1}$ (defining the ratio of the gradient of phase mismatch and absorption) will define the maximum flux (assuming no limit on medium length), resulting in similar optimization condition ($R_{1}^2 < 1$) to reach at least $50\%$ of the absorption-limited flux. This behaviour is depicted in Figure \ref{fig:rho0zero}(a). 

\begin{figure}[htb!]
	\begin{center}
		\includegraphics[width=16cm]{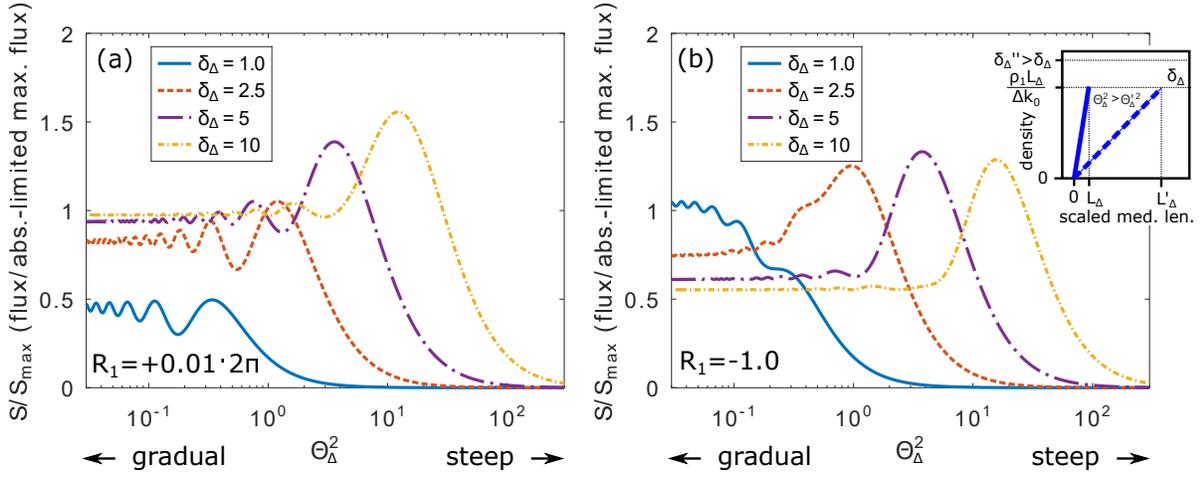}
	\end{center}
	\caption{\label{fig:fixedpress_rho0zero}The harmonic flux $S$ (normalized to the absorption-limited maximum $S_{\mathrm{max}}$ of Eq. (\ref{eq:Smax})) as a function of pressure increase steepness $\Theta_{\Delta}^{2} = \kappa_{1}/\Delta k_{0}^{2}$ for different peak pressures $\delta_{\Delta} = \rho_{\mathrm{peak}} \sigma/(2 \Delta k_{0})$. The medium has a linearly increasing pressure profile along the laser propagation direction (see inset of (b)). (a) For a phase mismatch increase with pressure ($R_{1} > 0$) and (b) for a phase mismatch decrease with pressure ($R_{1} < 0$).}
\end{figure}

The effect of pressure gradient (quantified by $\Theta_{\Delta}^{2}$) is studied in Figure \ref{fig:rho0zero}(b). One observation is that the gradient defines how fast the $S_{\infty}$ limit is reached upon propagation: higher gradient leads to shorter required medium length. 
Altogether, the optimizing conditions for a medium with linearly increasing pressure is very similar to those of a constant-pressure medium:
\begin{eqnarray}
	R_{1}^{2} < 1\,, \\
	L_{\Delta} > 1.5 \quad \mathrm{and} \\
	\Theta_{\Delta}^{2} > 1
\end{eqnarray}
guarantees that at least half of the absorption-limited maximum flux is achieved. For smaller gradients $\Theta_{\Delta}^2 \leq 1$, a longer medium length $L_{\Delta}$ is necessary. Also, for very steep gradient ($\Theta_{\Delta}^2 \gg 1$), it is physically not possible to reach the necessary medium lengths (see in relation Fig. \ref{fig:fixedpress_rho0zero}). A smoother gradient is typical for HHG gas cells, where the atoms exit the cell's volume through the same holes where the laser propagates through.

Another observation from Figure \ref{fig:rho0zero} is that with increasing gradient (increasing value of $\Theta_{\Delta}^{2}$) the flux at certain medium lengths can exceed the absorption-limited value at infinite medium length ($S_{\mathrm{max}}$, $1$ on the vertical axis of Fig. \ref{fig:rho0zero}(b)). This can be explained by the following.
The ratio between phase mismatch and absorption as a function of medium length is
\begin{equation}\label{eq:ratio}
	\frac{\Delta k}{\kappa} = R_{1} + \frac{1}{\Theta_{\Delta}^2 L_{\Delta}}\,.
\end{equation}
In case of a very long medium ($L_{\Delta} \rightarrow \infty$), this ratio reaches $R_{1}$, defining the infinite-medium-length, absorption-limited flux ($S_{\infty}$ of Eq. (\ref{eq:lim_infty}), with $R_{0}$ replaced by $R_{1}$, see details in \ref{Asec:first}). With higher density gradient $\Theta_{\Delta}^2$ ($>0$), the $\Delta k/\kappa$ ratio becomes smaller on a shorter propagation length, which is favorable for harmonic build up. Of course, at the same time $R_{1}$ value is also reached faster, setting earlier the absorption-limited flux. A negative value of $R_{1}$ can give even more favorable conditions: it means increasing phase matching ($\Delta k$ decreases) with almost no absorption along propagation, making curves run even closer to the absorption-free perfectly-phase-matched cases (black dashed curves in Figure \ref{fig:rho0zero}) in short propagation lengths. A situation describable by an increasing absorption and a decreasing phase mismatch at a similar rate (meaning $R_{1} \approx -1$, achievable only with low ionization levels of the medium according to considerations in \ref{Asec:dk_values}), gives typically better results for a lower pressure gradient (see Figures \ref{fig:rho0zero}(a) and \ref{fig:fixedpress_rho0zero}(b)). While analytical evaluation of the maximum is not possible because of the $\mathrm{erfi}(x)$ special function, numerical evaluation gives that the flux maximum cannot not exceed $\sim 1.65$-times the absorption limit with any gradient (see also Fig. \ref{fig:fixedpress_rho0zero}(a)).   

As an additional note, the absorption-free flux in this case evolves as $S_{\mathrm{abs-free}}/S_{\mathrm{max}} = \Theta_{\Delta}^4 L_{\Delta}^{4}/4$. This depends on the gradient $\Theta_{\Delta}^{2}$, so it cannot be represented as a single curve in Figure \ref{fig:rho0zero}(b), differently from the case of constant pressure (cf. Figure \ref{fig:Constant}). This is of simple reason: since absorption length changes along propagation, it has to be taken into account for the absorption-free case when having the horizontal axis in the dimensionless units of $L_{\Delta} = \Delta k_{0} L_{\mathrm{med}}$. 

To address the question of achievable flux from a more application-oriented viewpoint, we compare cases where the same peak pressures are reached with different pressure gradients in Fig. \ref{fig:fixedpress_rho0zero}. The density gradient can be tuned experimentally by changing the distance between the nozzle orifice and the generating laser beam \cite{Altucci1996JOSAB, Luria2011JPCA, Congsen2014RSI, Comby2018OE}, for example. In case of capillaries or gas cells, specially designed gas inlets and outlets, or entrance and exit holes, can provide the desired pressure gradients \cite{Goh2015OE}. 
It is easy to see that to reach a certain peak pressure $\rho_{\mathrm{peak}}$ with a density increase characterized by $\rho_{1}$, the necessary medium length is changing according to
\begin{equation}\label{eq:delta}
L_{\Delta, \mathrm{peak}} = \delta_{\Delta}/\Theta_{\Delta}^{2}\,,
\end{equation}
where $\delta_{\Delta} = \rho_{\mathrm{peak}} \sigma/(2 \Delta k_{0})$ is a dimensionless measure of peak pressure. Following the considerations in \ref{Asec:dk_values}, $\delta_{\Delta}$ typically ranges between $\sim 0.1$ and $\sim 10$ for usual HHG conditions. Actually, relevant changes in the steepness dependence of harmonic flux (Figure \ref{fig:fixedpress_rho0zero}) only happens when $1 < \delta_{\Delta} < 10$, smaller (higher) values leading qualitatively to only a horizontal shift of the respective curves in Fig. \ref{fig:fixedpress_rho0zero} towards more gradual (steeper) gradients accompanied by an amplitude decrease. 

The main conclusions of Fig. \ref{fig:fixedpress_rho0zero} in addition to those drawn from Fig. \ref{fig:rho0zero} are the following. As can be seen in both sub-figures of Fig. \ref{fig:fixedpress_rho0zero}, for higher peak pressures the optimum gradient steepness resulting the highest flux is bigger. In line with Eq. (\ref{eq:rho0zero}), for decreasing steepness the harmonic flux is the absorption limited maximum (see Eq. (\ref{eq:lim_infty})) , since the medium length $L_{\Delta, \mathrm{peak}} \rightarrow \infty$ with $\Theta_{\Delta}^{2} \rightarrow 0$. Oppositely, with increasing steepness ($\Theta_{\Delta}^{2}$ increasing) the flux tends to zero, since $L_{\Delta, \mathrm{peak}} \rightarrow 0$, and there is not enough propagation length for harmonic field build up. As a result, there is an optimum density gradient to reach the highest harmonic flux for a certain peak pressure, typically ranging between $\Theta_{\Delta}^2 = 1$ and $10$.
What pressure increase this $\Theta_{\Delta}^2$ values means physically depends on the coherence length, defined by several generation conditions (see Section \ref{subsec:dk} and the definition of $\Theta_{\Delta}^2$).  
An interesting conclusion from Fig. \ref{fig:fixedpress_rho0zero}(b) is that for low ionization levels (few percent ionization rate necessary for $R_{1}<0$, see \ref{Asec:dk_values}) a long, gradual increase of pressure is an alternative solution for optimized flux (cf. the blue continuous curve with the others at gradual values in Fig. \ref{fig:fixedpress_rho0zero}(b)).

\subsection{Phase matching in medium with linearly decreasing pressure}\label{subsec:rho0nonzero}

\begin{figure}[htb!]
	\begin{center}
		\includegraphics[width=16cm]{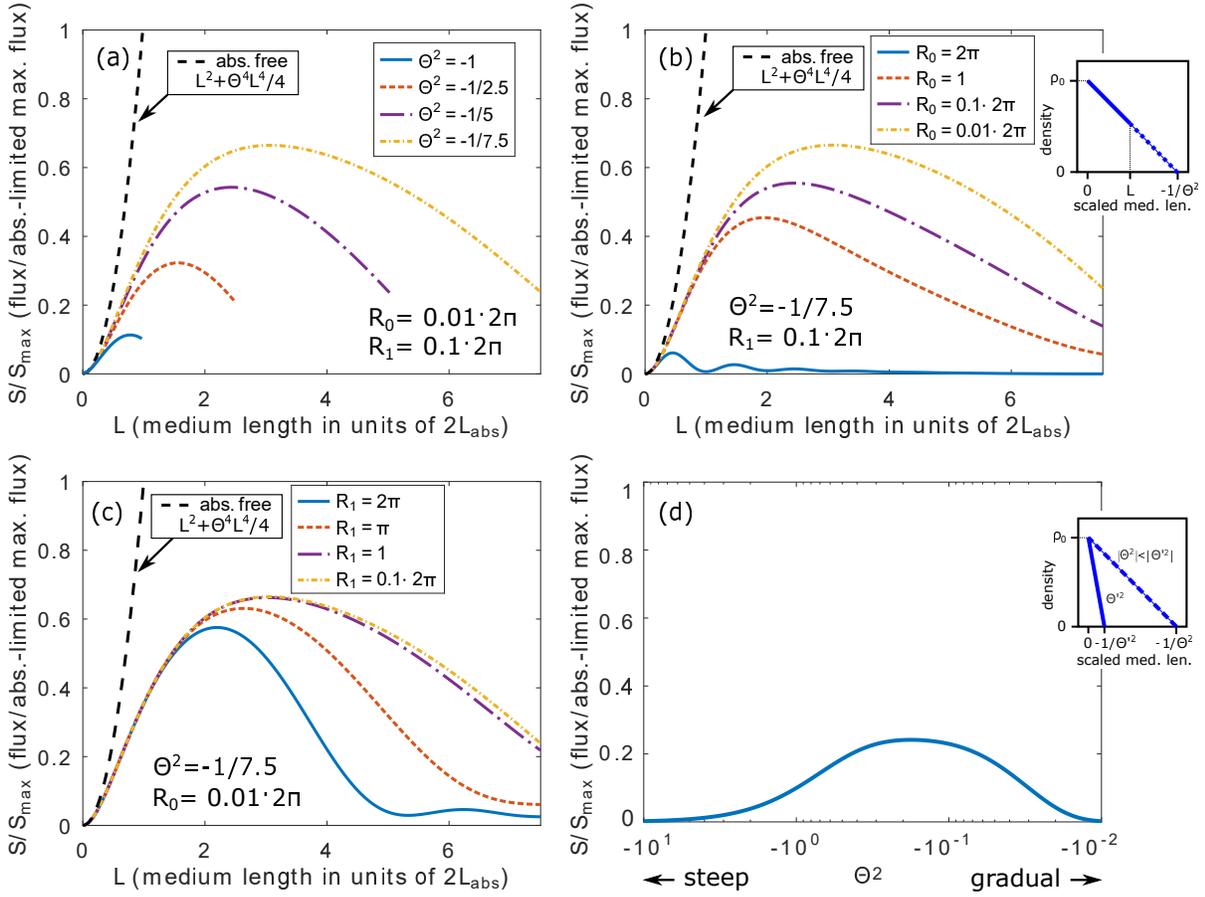}
	\end{center}
	\caption{\label{fig:rho0nonzero} (a), (b) and (c) The harmonic flux $S$ (normalized to the absorption-limited maximum $S_{\mathrm{max}}$ of Eq. (\ref{eq:Smax})) as a function of scaled medium length $L$. The medium has a linearly decreasing pressure profile along the laser propagation direction (see inset in (b)). (a) Dependence of flux on gradient steepness quantified by $\Theta^2$. (b) The same dependence on ratio $R_{0}$ between the phase mismatch and absorption at $z=0$. (c) High-harmonic flux for different ratios $R_{1}$ of phase-mismatch and absorption gradients. The black dashed curves show the absorption-free, perfectly phase matched case, depending on the value of $\Theta^{2}$.
	(d) The harmonic flux as a function of the pressure decrease steepness for a fixed beginning density $\rho_{0}$ (see inset of (d)). }
\end{figure}

Now consider a pressure profile of the form $\rho(z) = \rho_{0} + \rho_{1} z$ with $ \rho_{1} < 0$ (see inset of Figure \ref{fig:rho0nonzero}). As a result, absorption is also varying as $\kappa(z) = \kappa_{0} + \kappa_{1} z$ (with $\kappa_{1} < 0$). This is the simplest form to consider the end gradient of a medium along the generation laser beam propagation axis. The phase mismatch is assumed to be varying as $\Delta k(z) = \Delta k_{0} + \Delta k_{1} z$. Considerations on this form of phase mismatch are discussed in Section \ref{subsec:dk}.

With the above parameters, the integral of Eq. (\ref{eq:integral}) can be evaluated to be (see details in \ref{Asec:first}):

\begin{eqnarray}\label{eq:rho0nonzero}
	E_{q} = \frac{A}{\sigma} \frac{1}{1- i R_{1}} \times \nonumber \\ 
	\left\{ \sqrt{\pi} \left[ f + g(1-h) \right] 
	\exp{\left( - f^{2} \right)} \right. 
	\left[ \mathrm{erfi} \left( f \right)
	+ \mathrm{erfi} \left(  f - g h \right)  \right] \nonumber \\
	+ \left. \left[ 1 - \exp\left( g^2 h^2 - 2 g h f \right)  \right] \right\}
\end{eqnarray}
with dependence on medium length $L = \kappa_{0} L_{\mathrm{med}}$ only through
\begin{equation}\label{eq:f_rho0nonzero}
	f(L) = \frac{g}{2} \left[ \Theta^{2} L + h \right],\,
\end{equation}
and $\Theta^{2} = \kappa_{1}/\kappa_{0}^{2}$ ($<0$), $g = \sqrt{1-iR_{1}}/\Theta $, $R_{0} = \Delta k_{0}/\kappa_{0}$, $R_{1} = \Delta k_{1}/\kappa_{1}$, $h = (1-iR_{0})/(1-iR_{1})$. This is formally very similar to Eq. (\ref{eq:rho0zero}), but the variables are identical to that of the zero-order formula of Section \ref{sec:Methods}. Compared to the case of a pressure profile with linear increase, there is one extra dimensionless variable ($R_{1}$ and $R_{0}$ are both present), increasing their number to four: $L$, $R_{0}$, $R_{1}$ and $\Theta^2$. 

The evolution of harmonic flux in a medium with linearly decreasing pressure as a function of propagation distance is analyzed in Figure \ref{fig:rho0nonzero}(a)-(c). The first and most important observation is that the medium length in this case is limited to $L=-1/\Theta^2$ (see Figure \ref{fig:rho0nonzero}(a)), since beyond this value the pressure/density would become negative and non-physical. Just like in the case of increasing pressure, the absorption-free case is $\Theta^{2}$ dependent, given by $S_{\mathrm{abs-free}}/S_{\mathrm{max}} = L^2 + \Theta^4 L^{4}/4$ (see black dashed curves in Figure \ref{fig:rho0nonzero}).

Actually, in realistic situations the medium length is always exactly $L=-1/\Theta^2$, since the pressure always reaches zero with a gradual decrease. So the fluxes depicted in Fig. \ref{fig:rho0nonzero}(a)-(c) with medium lengths less than $L=-1/\Theta^2$ are for pressure profiles of trapezion form (a trapezoid with two parallel vertical sides, see the blue continuous curve in the inset of Fig. \ref{fig:rho0nonzero}(b)).
The key observation here is that with the assumed negative gradient of pressure, harmonic flux never reaches the absorption-limited maximum $S_{\mathrm{max}}$ ($1$ on the vertical axes of Figure \ref{fig:rho0nonzero}). Instead, evaluating Eq. (\ref{eq:rho0nonzero}) at $L=-1/\Theta^2$ leads to an expression with a maximum of $A^2/\sigma^2/(1+R_{1}^2)$ (see \ref{Asec:first} for details). This means that the achievable flux is limited to the $1/4$ of the absorption-limited maximum $S_{\mathrm{max}}$ for such pressure profile (see also Fig. \ref{fig:rho0nonzero}(d)). Situation $R_{1}<0$ is not analyzed, not having relevant differences from $R_{1}>0$ cases. 

In Fig. \ref{fig:rho0nonzero}(d) we study the effect of pressure decrease steepness on harmonic flux for a fixed beginning pressure $\rho_{0}$ (inset of Fig. \ref{fig:rho0nonzero}(d)), similarly to the last paragraph on the linearly increasing pressure case. The situation is simpler mathematically compared to Section \ref{subsec:rho0zero} and Fig. \ref{fig:fixedpress_rho0zero}, because a single curve describes the variation for any $\rho_{0}$ pressure assumed in the beginning of the medium. As written earlier, the medium length for a certain gradient in this case is given by the equation $L=-1/\Theta^2$.
As can be seen in Fig. \ref{fig:rho0nonzero}(d), the harmonic flux is limited to $25\%$ of the absorption-limited maximum $S_{\mathrm{max}}$. The optimum gradient for a triangle shaped pressure profile with step-like beginning is $\Theta^{2} \approx -0.2$, which means that the linear pressure decrease should have an extent of about $10 L_{\mathrm{abs}}$ (ten times the absorption length $L_{\mathrm{abs}}$ corresponding to the beginning pressure $\rho_{0}$).  

\subsection{Effect of pressure gradient at the end of the medium on high-harmonic flux}

\begin{figure}[htb!]
	\begin{center}
		\includegraphics[width=16cm]{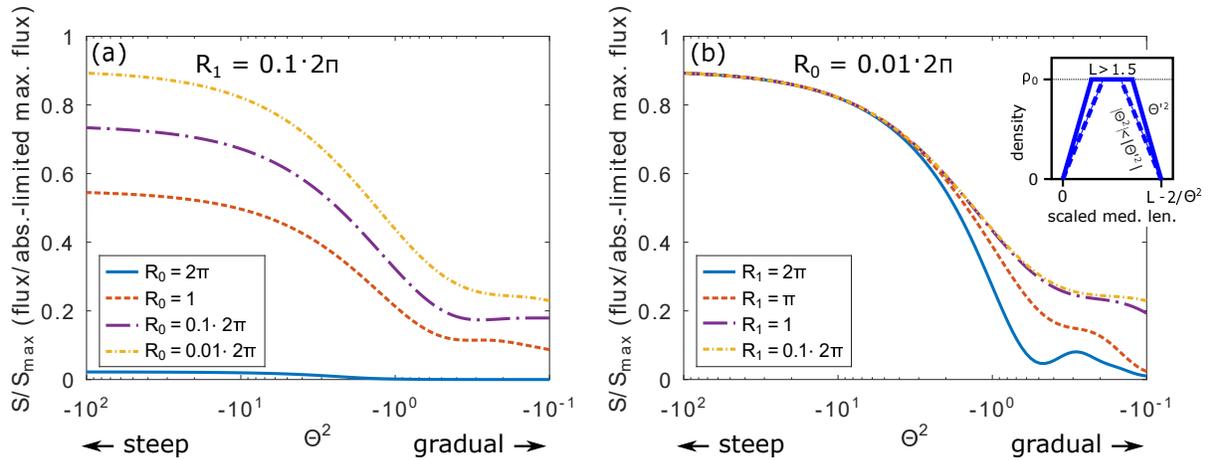}
	\end{center}
	\caption{\label{fig:endsteepness}The high-harmonic flux $S$ (normalized to the absorption-limited maximum $S_{\mathrm{max}}$ of Eq. (\ref{eq:Smax})) as a function of the pressure-gradient steepness $\Theta^2$ in case of a symmetric trapezoidal pressure profile (a) for different coherence and absorption length ratios $R_{0}$ at the constant pressure profile region, (b) for different ratios $R_{1}$ of phase mismatch and absorption gradients in the gradient pressure region. The inset in (b) shows the assumed pressure profile. As indicated in the inset, in case of all gradients the constant region extends $L>1.5$, guaranteeing long enough medium for harmonic build up, as suggested by Sections \ref{sec:Methods} and \ref{subsec:rho0zero}.}
\end{figure}

After the results of the preceding section, the question arises on the effect of pressure gradient at the end of medium on the achievable high-harmonic flux. To analyze this we consider a trapezoidal pressure profile with the same gradients both at the beginning and at the end (see inset of Figure \ref{fig:endsteepness}). Such symmetric pressure profile serves as a fair approximation of realistic generation media when using gas jets \cite{Luria2011JPCA, Congsen2014RSI, Altucci1996JOSAB, Comby2018OE} or cells \cite{Major2020JPysPhot}. While in this case $E_{q}$ still can be evaluated analytically, resulting in a closed-form expression, numerical evaluation of Eq. (\ref{eq:integral}) has been carried out due to the complexity of the analytical expression. The presented results, however, still hold in general, since as concluded in previous sections (Sections \ref{sec:Methods} and \ref{subsec:rho0zero}), if one applies the general optimization rules of Constant \emph{et al.} \cite{Constant1999PRL} --- meaning that a longer coherence length is maintained in the medium than the absorption length, and the medium length is long enough --- the absorption-limited flux ($S_{\infty}$ of Eq. (\ref{eq:lim_infty})) can be reached. To assure that the absorption-limited flux is built up in the medium, length of the constant-pressure region of the trapezoidal profile is always $L>1.5$ (see Figure \ref{fig:endsteepness}). Also, we have tested the effect of medium-end pressure-steepness on different simplified pressure profiles like triangular or a constant profile followed by a gradual decrease (considering the requirement $L>1.5$ on medium length to achieve the absorption-limited flux), and all lead to an identical conclusion to what is depicted in Figure \ref{fig:endsteepness}: a steep pressure drop is critical for not loosing flux at the end of the medium.

As can be seen in Figure \ref{fig:endsteepness}(a), if $\Theta^{2} < -10$ approximately $4/5$ the absorption limited flux can be achieved for a certain $R_{0}$, while a gradual decrease of $\Theta^{2} > -1$ gives only $\sim 2/5$. As a better guideline for experiments --- using that $\rho = 0$ if $L = -1/\Theta^2$ ---, the requirement on $\Theta^{2}$ gives that for keeping at least $80\%$ of the flux, pressure drop has to have a steepness that guarantees zero pressure within $1/5$ of the absorption length (defined by the density in the constant region). If the range of linear pressure drop is longer than twice the absorption length, only $\sim 40\%$ of the flux remains. Figure \ref{fig:endsteepness}(b)  highlights the importance for steep pressure decrease from a different aspect. The achievable flux is independent of the ratio $R_{1}$ of phase mismatch and absorption gradients (see overlapping curves of \ref{fig:endsteepness}(b) when $\Theta^{2} < -10$). The required steepness increases (meaning larger modulus of $\Theta^2$) with increasing $R_{1}$. 

Together with findings in Section \ref{subsec:rho0zero}, this simple one-dimensional model suggests that the ideal medium for high flux is a volume of atoms with certain pressure gradient in the beginning and an abrupt ending (e.g., a gas jet with special orifice design). At the same time, a more gradual pressure increase at few percent ionization rate  can also give similarly high flux, but in this case a longer propagation length is required (e.g., a gas-cell-based solution).

\section{Summary}\label{sec:Summary}

We have developed a first-order one-dimensional model for studying phase matching of high-harmonic generation in dispersive and absorptive medium. Thanks to the dimensionless form, general, universally applicable laws have been derived from the expressions presented in this work. We used these formulas to demonstrate that if the generating laser beam enters a generation medium with steep pressure/density gradient, a shorter medium is enough to reach the absorption-limited flux, and in favorable phase-matching conditions it can even lead to increased flux compared to what is achievable in a constant-pressure medium. In case of gas cells, which typically have a more gradual pressure increase along the laser propagation axis, a longer medium length is preferable to reach same flux. The formulas also suggest that while a certain, steep pressure gradient at the medium beginning (typically achievable with gas jets) will give an increased photon flux, at low ionization rates gas-jet and gas-cell-based approaches will behave similarly in terms of radiation strength.  We have also shown that if the laser beam exits the medium through a pressure gradient, it will always result in the decrease of harmonic yield, and only an abrupt drop of the density can guarantee that the achieved flux is maintained for the application.  

\section{Acknowledgment}
The ELI-ALPS project (GINOP-2.3.6-15-2015-00001) is supported by the European Union and co-financed by the European Regional Development Fund.

\appendix

\section{Typical values of phase mismatch and absorption in the HHG process}\label{Asec:dk_values}

In the following, typical value ranges of phase mismatch terms $\Delta k$ of Eq. (\ref{eq:dk4}) are summarized, along with usual absorption strengths $\kappa$. In this section it is assumed for all estimations that the studied harmonics are in the $20-200\,\mathrm{eV}$ photon energy range, the generating field has a wavelength of $500-3000\,\mathrm{nm}$, and the medium pressure is between $0$ and $2000\,\mathrm{mbar}$. The generation media studied are typical noble gases: Xe, Kr, Ar and Ne.
  
The values of geometrical phase mismatch $\Delta k_{g}$ for a non-guiding geometry is usually a fraction of the focused fundamental beam's Rayleigh length \cite{Heyl2016JPB}, so takes typical values between $\Delta k_{g} \approx - 10^{-7}$ and $- 10^{-4}\,\mathrm{1/nm}$ (assuming typical Rayleigh lengths ranging from sub-millimeter to centimeters, and harmonic orders on the order of a few tens or hundreds). For guiding geometries, a much higher degree of freedom is available \cite{Heyl2016JPB, Durfee1999PRL}.

It can be shown by using tabulated values of the refractive indeces of noble gases in the infrared \cite{Bideau1981JQSRT, Borzsonyi2008AO} and XUV wavelength ranges \cite{Henke1993ADNDT} --- or by evaluating Eq. (\ref{eq:dkn}) --- that in the analyzed photon energy range the coefficient of Eq. (\ref{eq:dkn}) is typically $\gamma_{n} \approx 0- 10^{-2}\,\mathrm{nm^2}$. In the pressure range of $0-2\,\mathrm{bar}$ the atomic number density is $\rho \approx 0-10^{-1}\,\mathrm{1/nm^3}$, leading to neutral phase mismatch of $\Delta k_{n} \approx 0 - 10^{-3}\,\mathrm{1/nm}$ according to Eq. (\ref{eq:dkn}).

The value of the phase mismatch coefficient $\gamma_{p}$ in Eq. (\ref{eq:dkp}) is independent of the gas type, and can be obtained to between $-0.5$ and $-0.05\,\mathrm{nm^2}$ with $100\%$ ionization rate. The fact that the magnitude of $\gamma_{p}$ is at least an order of magnitude larger than $\gamma_{n}$ results in the often described property that phase matching is achievable only with a few percent ionization rate \cite{Rudawski2013RSI, Heyl2016JPB}. Accordingly, the free-electron-induced phase mismatch ranges between $\Delta k_{p} \approx -5\cdot 10^{-2}\,\mathrm{1/nm}$ and zero, depending on the ionization rate.

The absorption of high-order harmonics can also be obtained using measured and simulated photoionization cross-sections \cite{Henke1993ADNDT, Krebs2014AJP}, typically ranging between $\sigma \approx 10^{-3} - 10^{-4}\,\mathrm{nm^2}$. These correspond to absorption strengths of $\kappa \approx 10^{-7} - 10^{-5}\,\mathrm{1/nm}$ for few hundreds of millibar pressure. 

From the above values it is easy to see using Eqs. (\ref{eq:dk0}) and (\ref{eq:dk1}) that the phase mismatch coefficients $\Delta k_0$ and $\Delta k_1$ under usual HHG conditions can vary between $-10^{-2}$ and $+10^{-2}\,\mathrm{1/nm}$. These result in values for the dimensionless quantities of Sections \ref{sec:Methods} and \ref{sec:Results} that show a broader range than what is plotted in relevant figures. Plot ranges were chosen instead based on what parameter values give relevant changes in the results.    

\section{Zeroth-order, one-dimensional model for phase matching}\label{Asec:zeroth}

To obtain the flux $S_{q}$ of high harmonics on axis in a medium with constant properties, one has to evaluate the integral of Eq. (\ref{eq:integral}), giving 
\begin{eqnarray}
	E_{q} =&  \int_{0}^{L_{\mathrm{med}}}  A \rho_{0}
	\exp \left( \mathrm{i} [\Delta k_{0} + \mathrm{i} \kappa_{0}]
	[L_{\mathrm{med}}-z] \right) \mathrm{d}z  = \nonumber \\
	 &A \rho_{0} \left[ \frac{\exp \left( \mathrm{i} [\Delta k_{0} + \mathrm{i} \kappa_{0}]
		[L_{\mathrm{med}}-z]\right)}{-i [\Delta k_{0} + i \kappa_{0}]} \right]_{0}^{L_{\mathrm{med}}} \,,
\end{eqnarray}
using the Newton-Leibniz axiom. After some algebra, the modulus square $S_{q} = |E_{q}|^2$ can be obtained to be \cite{Heyl2016JPB}\footnote{Note that a factor $2$ is missing in Eq.~(11) of Ref. \cite{Heyl2016JPB}.} 

\begin{equation}\label{Aeq:Heyl}
	S_{q} = 2  A_{q}^{2} \rho_{0}^{2} \exp\left(-\kappa_{0} L_{\mathrm{med}}\right)
	\frac{\cosh (\kappa_{0} L_{\mathrm{med}}) - \cos(\Delta k_{0} L_{\mathrm{med}})}{\Delta k_{0}^{2} 
		+ \kappa_{0}^{2}}\,.
\end{equation}

Using that the coherence length is related to the wave vector mismatch of harmonic order $q$ according to  $L_{\mathrm{coh}} = \pi / \Delta k$, while the absorption length is defined as $L_{\mathrm{abs}} = 1 / (2 \kappa_{0})$, and that $\mathrm{cosh}(x) = [1 + \exp(-2x)]/[2\exp(-x)]$, one can modify Eq. (\ref{Aeq:Heyl}) to obtain the equivalent expression of Constant~\emph{et. al.} \cite{Constant1999PRL}

\begin{eqnarray}
	S_{q} = &  \rho_{0}^{2} A_{q}^{2} \frac{4 L_{\mathrm{abs}}^{2}}{1 + 
		4 \pi^{2}L_{\mathrm{abs}}^{2} / L_{\mathrm{coh}}^{2}} \times \nonumber \\
	& \left[
	1 + \exp \left( - \frac{L_{\mathrm{med}}}{L_{\mathrm{abs}}} \right)
	- 2 \cos \left( \frac{\pi L_{\mathrm{med}}}{L_{\mathrm{coh}}} \right)
	\exp \left( -\frac{L_{\mathrm{med}}}{2L_{\mathrm{abs}}} \right)
	\right]\,.
\end{eqnarray}

Using the relation $\rho_{0} = 2\kappa_{0}/\sigma = 1/(\sigma L_{\mathrm{abs}})$ ($\sigma$ being the photoionization cross section) and introducing the dimensionless variables of Section \ref{sec:Methods} results in the expressions (\ref{eq:Heyl_dimless}) and (\ref{eq:Constant_dimless}).

\section{First-order, one-dimensional model for phase matching}\label{Asec:first}

In case of the first-order expression (see Section \ref{sec:Results}), the integral needing evaluation has an explicit form of 
\begin{eqnarray}
	E_{q} =   \int_{0}^{L_{\mathrm{med}}}  A& [\rho_{0} + \rho_{1}z] \times \nonumber \\
	&\exp \left( i \left[\Delta k_{0} + \Delta k_{1}z + i (\kappa_{0} + \kappa_{1}z)\right]
	\left[L_{\mathrm{med}}-z\right] \right) \mathrm{d}z \,.
\end{eqnarray}

The expression above can be written equivalently in the form
\begin{eqnarray} \label{Aeq:int_terms}
	E_{q} = & A \rho_{0} \exp\left([-\kappa_{0} + i\Delta k_{0}] L_{\mathrm{med}} \right) \int_{0}^{L_{\mathrm{med}}} \exp \left( a z^2 + bz \right) \mathrm{d}z + \nonumber \\
	&A \rho_{1} \exp\left([-\kappa_{0} + i\Delta k_{0}] L_{\mathrm{med}} \right) \int_{0}^{L_{\mathrm{med}}} z \exp \left( a z^2 + b z \right) \mathrm{d}z \,,
\end{eqnarray}
with $a = \kappa_{1} -i \Delta k_{1} = k_{1}$ and $b = \kappa_{0} - i \Delta k_{0}- \kappa_{1} L_{\mathrm{med}} + i \Delta k_{1} L_{\mathrm{med}} = k_{0} - k_{1} L_{\mathrm{med}}$. These integrals can be evaluated analytically using expressions 2.325.13 and 3.321.4 of Ref. \cite{gradshteyn2007}, which with slight formal modifications read as
\begin{equation}\label{Aeq:GR1}
	\int  \exp \left( a x^2 + bx \right) \mathrm{d}x = \frac{1}{2}\sqrt\frac{\pi}{a} \exp\left(-\frac{b^2}{4a} \right) \mathrm{erfi} \left( \frac{2 ax + b}{2 \sqrt{a}}\right) \quad (a \neq 0)
\end{equation} 
and
\begin{equation}\label{Aeq:GR2}
	\int_{0}^{u} x \exp(-v^2 x^2)\,\mathrm{d}x = \frac{1}{2v^2} \left[ 1 - \exp(-v^2 u^2) \right]\,,
\end{equation} 
where 
\begin{equation}\label{Aeq:erfi}
	\mathrm{erfi}(x) = -i \mathrm{erf}(ix) = \frac{2}{\sqrt{\pi}} \int_{0}^{x} \exp(t^2)\,\mathrm{d}t
\end{equation} 
is the imaginary error function \cite{gradshteyn2007}.
The first term in Eq. (\ref{Aeq:int_terms}) can be directly evaluated using the tabulated integral of Eq. (\ref{Aeq:GR1}).  After completing the square in the exponential of the second term in Eq. (\ref{Aeq:int_terms}), it can be written as the sum of terms formally equivalent to Eqs. (\ref{Aeq:GR1}) and (\ref{Aeq:GR2}).
This way one obtains
\begin{eqnarray} \label{Aeq:simplified}
	E_{q} = & A 
	 \left[ \rho_{0} + \left( \frac{L_{\mathrm{med}}}{2} - \frac{k_{0}}{2 k_1}  \right)\rho_{1}   \right] \times \nonumber \\
	&\sqrt\frac{\pi}{4k_1} \exp\left(-\left[ \frac{\sqrt{k_1}L_{\mathrm{med}} }{2} + \frac{k_0}{2\sqrt{k_1}} \right]^{2} \right) \times  \\ 
	&\left[ \mathrm{erfi} \left( \frac{\sqrt{k_1}L_{\mathrm{med}} }{2} + \frac{k_0}{2\sqrt{k_1}}\right) - \mathrm{erfi} \left(  -\frac{\sqrt{k_1}L_{\mathrm{med}} }{2} + \frac{k_0}{2\sqrt{k_1}}\right)  \right] +
    \nonumber \\
	& A\frac{\rho_{1}}{2k_1} \left[ 1- \exp(-k_0 L_{\mathrm{med}} \right] \,.
\end{eqnarray}
By using the relation $L_{\mathrm{abs}} = 1/(\rho \sigma)$ \cite{Constant1999PRL}$ ( \Rightarrow \rho_{0} = (2/\sigma) \kappa_{0}, \, \rho_{1} = (2/\sigma) \kappa_{1}$) to make the results independent of pressure, and after some algebra, one can obtain both Eq. (\ref{eq:rho0zero}) and Eq. (\ref{eq:rho0nonzero}) after introducing the necessary dimensionless variables of Sections \ref{subsec:rho0zero} and \ref{subsec:rho0nonzero}, respectively.  

The limit in $L_{\mathrm{med}} \rightarrow \infty$ of Eq.  (\ref{eq:rho0zero}) can be evaluated using the series expansion of the imaginary error function around $x=\infty$ \cite{abramowitzstegun}
\begin{equation}
	\mathrm{erfi}(x+c) = -i + \exp([x+c]^2) \left[ \frac{1}{\sqrt{\pi}x} + O\left( \frac{1}{x^2}\right) \right]\,,
\end{equation}
leading to $2$ for the expression in curly brackets of Eq. (\ref{eq:rho0zero}).

The maximum of Eq. (\ref{eq:rho0nonzero}) can be found after the following considerations. As is shown by the curves of Figure \ref{fig:rho0nonzero}, and also suggested by physical considerations, highest flux is achievable if phase mismatch is much smaller than absorption, mathematically meaning that $R_{0} \rightarrow 0$ and $R_{1}  \rightarrow 0$. As a consequence, $h = (1-iR_{0})/(1-iR_{1}) \rightarrow 1$. With these parameters evaluating $f(L)$ (see Eq. (\ref{eq:f_rho0nonzero})) at the medium end ($L = -1/\Theta^2$) leads to $f = 0$. As a result, Eq. (\ref{eq:rho0nonzero}) gives
\begin{equation}\label{Aeq:max_rho0nonzero}
	E_{q} = \frac{A}{\sigma} \frac{1}{1- i R_{1}} \left[ 1 - \exp(g^2)\right]\,.
\end{equation}
Considering that $\Theta^2 < 0 \Rightarrow \mathrm{Re}[g^2] < 0$, the expression in the rectangular brackets of Eq. (\ref{Aeq:max_rho0nonzero}) has a maximum of $1$.

\section*{References}
\providecommand{\newblock}{}

\bibliographystyle{iopart-num}

\end{document}